\documentstyle[fleqn,espcrc2,epsfig]{article}

\title{Latest Results on Weak Semi-Leptonic Hyperon Decays in KTeV} 
\author {Nickolas Solomey \address {Enrico 
Fermi Institute, The University of Chicago, 
Chicago, IL 60637, USA}}

\begin{document}

\begin{abstract}
The KTeV experiment at Fermilab has finished its $\Xi^0$ semi-leptonic
decay analysis from the 1997 run and presented here are the final
numbers for the branching ratio and form factors. A new run with
three times the statistics was taken in 1999 and a first look at the
data shows the quality and consistency of the results, predicting 
a similar improvement in these measurements.
\end{abstract}

\maketitle

\section{Semi-Leptonic $\Xi^0$ Decays}
Semi-leptonic hyperon decays are important as
tests for new physics by measuring their form factors and branching
ratios \cite{theory1}. This work started experimentally with 
$\Lambda$ beta decay \cite{Jenson,Earl}:
\begin{equation}
\Lambda \rightarrow p e^- \bar\nu.
\end{equation}
It was followed by a high statistics sample of $\Sigma^-$ beta 
decays \cite{E715}:
\begin{equation}
\Sigma^- \rightarrow n e^- \bar\nu,
\end{equation}
and $\Xi^-$ beta decays \cite{WA2}:
\begin{equation}
\Xi^- \rightarrow \Lambda e^- \bar\nu  \:\:\:\:\:  and  \:\:\:\:\:
\Xi^- \rightarrow \Sigma^0 e^- \bar\nu.
\end{equation}

It was proposed in 1993 to extend the KTeV program \cite{ktev} to include a
search for the unseen $\Xi^0$ beta decays \cite{Solomey}:
\begin{equation}
\Xi^0 \rightarrow \Sigma^+ e^- \bar\nu.
\end{equation}
This suggestion of the author was shown to be accessible in KTeV because
even though there was no charged particle vertex, the $\pi^0$ decay into
two gammas and the high precision electronmagnetic calorimeter
permits the vertex to be precisely measured by constraining to 
the $\pi^0$ mass \cite{Tony}. This along with the high flux of
$\Xi^0$ hyperons in the experiment permitted these analysis.

Hyperon semi-leptonic decays with a muon replacing the electron are
only energetically permitted for a few decay modes. 
In all there were only three types of such decays ever
observed and all of these with a very small number 
of events. It was the author who introduced
and performed the analysis in KTeV of the $\Xi^0$ muonic decay \cite{Solomey2}.
These are important because they can be a first glimpse at the g$_3$
form factor which is multiplied by (M$_{lepton}$/M$_{hyperon}$);
in beta decays this ratio squelches any effect of the g$_3$ form factor, but
in muonic decays this term, although small, is no longer negligible 
\cite{linke}. 

\section{Final Results from KTeV-97}
Many summary talks of the KTeV hyperon data and analysis techniques
are available elsewhere \cite{nick}. The goal of this section is
to summarize in one place the final numbers available at this time. 
A Physical Review Letter was put out with a fraction of
the KTeV 1997 data called the winter run, yielding a branching ratio 
based on 176 events of
2.71~($\pm$0.22~$\pm$0.31)~$\times$~10$^{-4}$ \cite{cbet},
where the first error is statistical and the second systematic.
Since then the remaining 1997 data sample has been analyzed 
\cite{wisc} and
the branching ratio for the $\Xi^0$ beta decay from this separate sample is
2.60~($\pm$0.11~$\pm$0.16)~$\times$~10$^{-4}$,
based on 626 events. The theoretical expected is 2.6~$\times$~10$^{-4}$.
The $\Xi^0$ muonic decay has a preliminary branching ratio of
3.5~($^{+2.0}_{-1.0}$~$^{+0.5}_{-1.0}$)~$\times$~10$^{-6}$ based on 5 events,
again the first error is statistical and the second systematic the
asymmetric errors are due to the small sample of events; the 
theoretical expected is 2.6~$\times$~10$^{-6}$.
 
\begin{figure}
\begin{center}
\mbox{
\epsfig{file=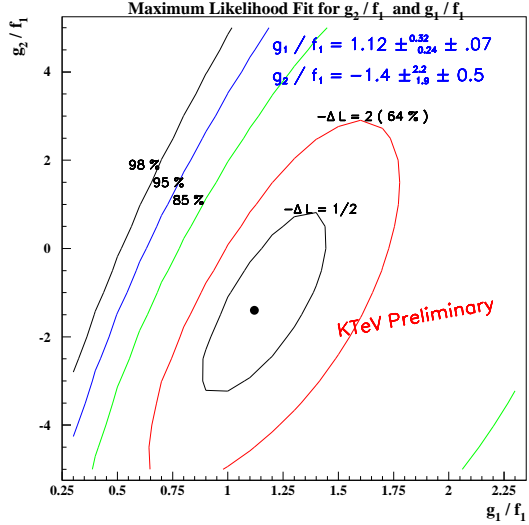,%
height=0.99\linewidth}
}
\end{center}
\caption{Maximum likelihood fit of the form factors g$_1$ and g$_2$,
the later is consistent with zero.}
\label{g2}
\end{figure}

The final four form factors obtained in our analysis \cite{bright} using a 
ultra clean sample of $\Xi^0$ beta decays by using the added power
of the TRD particle identification \cite{trd} are:
$f_1 = 0.99 \pm 0.14$, $\frac{g_1}{f_1}=1.12 \pm 0.27$,
$\frac{f_2}{f_1}=2.3 \pm 1.3$, and $\frac{g_2}{f_1}= -1.4 \pm 2.1$
(the errors are combined statistical and systematic for brevity);
here this analysis used the previously quoted branching ratio, and permitted
the $g_2$ form factor to float. The $g_2$ form factor is consistent with
zero, see figure \ref{g2},
and in another analysis it was constrained to be zero and the remaining three
form factors reanalyzed and they essentially remained unchanged.

\section{Improved Form Factors}
\begin{figure}
\begin{center}
\mbox{
\epsfig{file=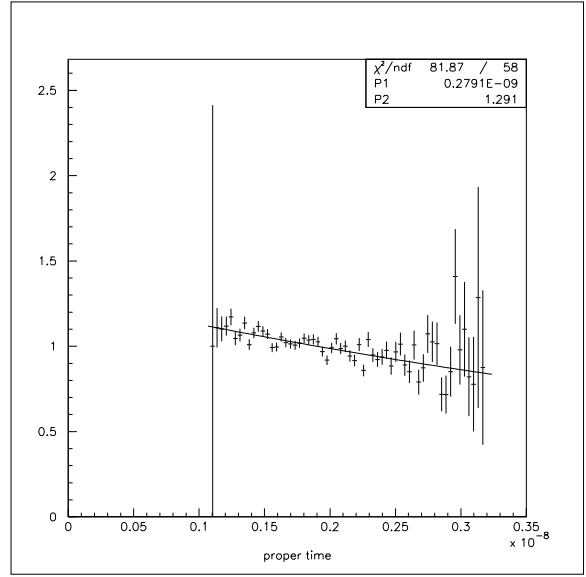,%
height=0.99\linewidth}
}
\end{center}
\caption{Ratio of data and normalized Monte-Carlo of the proper time (in
seconds) of the decay $\Xi^0 \rightarrow \Lambda^0 \pi^0$. Our
preliminary lifetime measurement is 2.79($\pm$0.01 $\pm$0.05)$\times$10$^{-10}$ 
seconds, the first error is statistical based on a fit to 94839 events, the
second is the systematic error from z and p$_{\Xi^0}$ dependency.}
\label{lifetime}
\end{figure}

The form factors reported in the previous section are also dependent upon 
the mass of the $\Xi^0$ particle and its life time. If either of these 
numbers are off then this can have an effect upon the form factor 
results. The values used were those from the 1998 PDG \cite{pdg}. 
The KTeV experiment has tried improving both of these values: a first 
attempt at improving the mass was made \cite{Marlon}, but another 
experiment \cite{NA48} has released an improved value before our 
measurement was finished and they have an error that is better than 
what we would have achieved so this analysis was not pursued; 
however, we have made a factor of three 
improvement in the life time of the $\Xi^0$ \cite{Pietryko}, see 
figure \ref{lifetime}. Reevaluation of the form factors with these 
two improved values will lead to slightly different values, 
but consistent with the error bars.

An accurate determination of the $\Xi^0$ mass is also important as a test of
SU(3) symmetry in the Coleman-Glashow mass relationship between the hyperons:
\begin{equation}
M_{\Xi^-} - M_{\Xi^0} = (M_p - M_n) + (M_{\Sigma^-} - M_{\Sigma^+})
\end{equation}
Using the new mass value, yet to be incorporated in the PDG, 1314.82$\pm$0.06, 
this equates to 6.49$\pm$0.14~=~6.73$\pm$0.08, a 
small SU(3) symmetry breaking less than 3.6\%. However, there is a 
hyperon beta decay that is yet unseen that is an even better test:
\begin{equation}
\Xi^- \rightarrow \Xi^0 e^- \bar\nu.
\end{equation}
This is such a powerful test of SU(3) symmetry breaking because there is 
only $\sim$6.49 MeV of energy being released in this beta decay, so by measuring 
the branching ratio this is the most accurate way to measure the mass 
difference between the charged and neutral $\Xi$ particles.
\begin{figure}[h]
\begin{center}
\mbox{
\epsfig{file=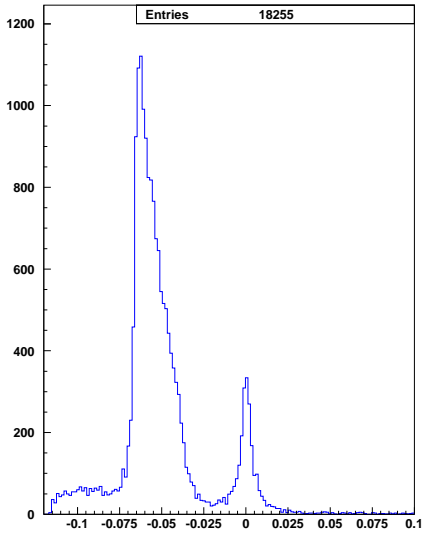,%
height=0.99\linewidth}
}
\end{center}
\caption{Preliminary results from the data obtained in
the KTeV run of 1999 using the analysis techniques developed for
the KTeV 1997 data sample \protect\cite{nick,wisc,bright} 
yields more than 2100 $\Xi^0$ beta decays
(small peak centered on zero), here
the horizontal axis is the mass of the reconstructed $\Sigma^+$ 
(i.e. p$\pi^0$ mass) subtracting its known value 1.189 GeV/c$^2$.
The peak on the left has been identified as several different 
background \protect\cite{bright}.}
\label{casbeta}
\end{figure}

\section{First results from KTeV-99}
\begin{figure}[h]
\begin{center}
\mbox{
\epsfig{file=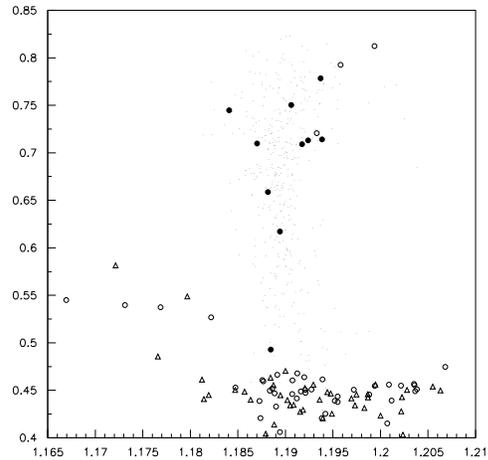,%
height=0.99\linewidth}
}
\end{center}
\caption{Preliminary results from the 
KTeV run in 1999 yields more $\Xi^0$ muonic decays. Here
both axis are in units of GeV/c$^2$,
on the horizontal is plotted the invariant mass of the
p$\pi^0$, and the vertical the mass ($\pi^+ \mu^- \pi^0$), 
from the dominant Kaon backgrounds:
either K$^0\rightarrow \pi^+ \pi^0 \pi^-$ where $\pi^- \rightarrow
\mu^- \bar\nu$ or K$\mu$4.}
\label{muonic}
\end{figure}

Data from a new run of KTeV in 1999 with improved triggers,
data acquisition and beam intensity is available for analysis. The
new run is expected to yield a three fold increase in $\Xi^0$ beta 
decays, see figure \ref{casbeta}, and a two 
fold increase in $\Xi^0$ muonic decays, see figure
\ref{muonic}.
The circle points are those events that satisfy the selection cuts
except for the mass analysis of the two axis and p$_t^2$ of the reconstructed
$\Xi^0$, and the triangle points are the opposite
charge sign analysis. There are far less anti $\Xi^0$ muonic decays 
expected in this wrong charge
sign analysis, due to a 10x suppression of anti $\Xi^0$ at production. 
The horizontal
band at the bottom are identifiable as the mentioned Kaon background decays, 
and the vertical band centered
on 1.189 GeV/c$^2$ are the $\Xi^0$ muonic decay events, 
the filled in data points are
those that satisfy the kinematic cuts from the 1997 blind analysis 
[p$_{t}^{2}<$0.0008~(GeV/c)$^2$ and various mass cuts].
The small dots are the signal Monte-Carlo for the final
event selection method. It
was decided that our blind analysis from the 1997 data was sufficiently
strong and straight forward enough that we did not need to make the 1999 
analysis blind, but it could be applied directly to the new data set.
This new KTeV data set should be calibrated by the end of 
2000 and improved branching ratio, form factors, lifetime and
possibly mass measurement within a year afterwards, depending
on manpower and computer resources.

\section{Conclusion}
The hyperon program in KTeV has 
observed both semi-leptonic decays of the $\Xi^{0}$. A high statistics
sample of the beta decay was used for 
measuring its branching ratio and form factors.
Furthermore, the observation of the muonic decay of the
neutral Cascade was observed with enough data 
to measure its branching ratio, but more data from the 1999 run is 
now available to improve this measurement. Furthermore, the $\Xi^0$
normal mode decay was also available to improve the mass and life-time 
values for this particle, important for SU(3) symmetry tests
as well as improving the form factor measurements.

\end{document}